\ifdefined\pdfminorversion
\fi
\documentclass[acmsmall,screen,nonacm]{acmart}

\usepackage{algorithmic}
\usepackage{algorithm}
\usepackage{graphicx}
\usepackage{textcomp}
\usepackage{xcolor}

\usepackage{makecell}
\usepackage{tabularx}
\usepackage[T1]{fontenc}
\usepackage{comment}
\usepackage{xcolor}
\usepackage{multirow}
\usepackage{array}
\usepackage{soul}
\usepackage{threeparttable}
\usepackage{balance}
\usepackage{diagbox}
\usepackage[normalem]{ulem}
\usepackage{subfigure}
\usepackage{microtype} 
\usepackage{calc}

\usepackage{enumitem}
\usepackage{url}
\usepackage{xspace}

\usepackage{float}
\usepackage{stfloats}
\usepackage{adjustbox}

\usepackage{amsmath,amssymb,amsfonts}
\usepackage{algorithmic}
\usepackage{graphicx}
\usepackage{textcomp}
\usepackage{xcolor}
\usepackage{multirow}

\usepackage[most]{tcolorbox}
\usepackage{makecell}
\usepackage{threeparttable}
\usepackage{colortbl}
\usepackage{subfigure}
\usepackage{hhline}
\usepackage{pifont}

\usepackage{enumitem}

\usepackage{booktabs}
\usepackage{siunitx}
\usepackage{booktabs}

\usepackage[utf8]{inputenc}
\usepackage[english]{babel}
\usepackage{enumitem}
\usepackage{bm}

\newcommand\app{RepoNorm\xspace}

\definecolor{reponormshade}{HTML}{F1F5F9}

\definecolor{evalincrease}{HTML}{21663A}
\definecolor{evaldecrease}{HTML}{A32935}
\newcommand{\evalup}[1]{\textcolor{evalincrease}{\(\uparrow\)#1}}
\newcommand{\evaldown}[1]{\textcolor{evaldecrease}{\(\downarrow\)#1}}
\newcommand{\evalflat}[1]{\textcolor{black!60}{#1}}

\definecolor{findingsaccent}{HTML}{334155}
\newtcolorbox{rqfindings}[1]{%
  enhanced,
  colback=reponormshade,
  coltext=black,
  frame hidden,
  borderline west={2pt}{0pt}{findingsaccent},
  sharp corners,
  boxrule=0pt,
  boxsep=0pt,
  left=8pt,right=8pt,top=3pt,bottom=3pt,
  before skip=\smallskipamount,
  after skip=\smallskipamount,
  fonttitle=\bfseries,
  coltitle=findingsaccent,
  title={RQ#1 Findings:},
  attach title to upper={\enspace},
}

\setcopyright{none}
\hypersetup{pdfauthor={Kaifeng He, Xiaojun Zhang, Zhenxi Chen, Christoph Treude, Mingwei Liu}}

\graphicspath{{imgs/}}

\begin{document}

\title{Acquiring and Verifying Repository Norms for Coding Agents}
\author{Kaifeng He}
\affiliation{%
  \institution{Sun Yat-Sen University}
  \country{China}}
\email{hekaifeng70@gmail.com}

\author{Xiaojun Zhang}
\affiliation{%
  \institution{Sun Yat-Sen University}
  \country{China}}
\email{zhangxj229@mail2.sysu.edu.cn}

\author{Zhenxi Chen}
\affiliation{%
  \institution{Sun Yat-Sen University}
  \country{China}}
\email{chenzhx236@mail2.sysu.edu.cn}

\author{Christoph Treude}
\affiliation{%
  \institution{Singapore Management University}
  \country{Singapore}}
\email{ctreude@smu.edu.sg}

\author{Mingwei Liu}
\authornote{Corresponding author.}
\affiliation{%
  \institution{Sun Yat-sen University}
  \country{China}}
\email{liumw26@mail.sysu.edu.cn}

\renewcommand{\shortauthors}{He et al.}

\begin{abstract}
Changes produced by coding agents can pass functional tests while leaving repository contribution requirements unmet. Following repository-specific norms requires identifying guidance dispersed across repository sources and interpreting its conditions and exceptions. Retrieval and documentation approaches supply general context, but agents must still determine which norms apply. We introduce \app to acquire explicit and implicit repository norms independently of coding tasks. It checks norm content and applicability using repository evidence, consults Git history when needed, and delivers norm packages to existing coding agents. Our evaluation uses three coding models and 121 tasks from RepoNormBench. Against the baseline with no additional generated guidance (Raw), relative improvements are 7.42--10.77\% for Overall Norm Compliance Rate (NCR), 31.64--45.44\% for Contribution NCR, and 11.34--17.69\% for Prompt-omitted NCR. All three coding models also obtain higher values on these NCR measures with \app than with CodeWiki documentation. Functional success rates show observed gains of 5.79--9.09 percentage points over Raw; the paired comparisons do not reach the significance threshold. Sampled precision is 86\% with \app's default configuration.
\end{abstract}

% Official LaTeX export from https://dl.acm.org/ccs,
% supplied by the author on 2026-09-29.
\begin{CCSXML}
<ccs2012>
   <concept>
       <concept_id>10011007.10011074.10011092</concept_id>
       <concept_desc>Software and its engineering~Software development techniques</concept_desc>
       <concept_significance>500</concept_significance>
       </concept>
 </ccs2012>
\end{CCSXML}

\ccsdesc[500]{Software and its engineering~Software development techniques}
\keywords{repository-specific norms, coding agents, norm acquisition,
  large language models, mining software repositories}

\maketitle

% Keep the class's default fonts, margins, line spacing, and heading styles.
\section{Introduction}
\label{sec:introduction}
Following established project practices is important for long-term software maintenance. Violating these practices can require corrective changes during code review~\cite{allamanis2014naturalize,sriiesaranusorn2021doescodereviewpromote}. \textbf{In this paper, we define repository-specific norms as requirements, recommendations, and supported source-code conventions that guide how changes should be implemented, tested, and documented in a project.} A norm identifies the changes or artifacts it governs, together with relevant conditions and exceptions. For example, Jinja's contribution guide explicitly requires a changelog entry with code changes. In Tiktoken, existing recovery code and its comment provide implicit evidence for preserving surrogate handling in new encoding paths. Coding agents now navigate repositories, edit files, and execute tests~\cite{yang2024sweagent,zhang2024autocoderoverautonomousprogramimprovement}. Their contributions must satisfy applicable requirements and respect other repository norms, even when task descriptions leave them unstated. \textbf{Overlooking norms can leave contributions incomplete or change established behavior despite passing functional tests, as illustrated in Section~\ref{sec:background}.}

One way to help agents follow these norms is through repository-level instruction files, such as \texttt{CLAUDE.md} or \texttt{AGENTS.md}~\cite{chatlatanagulchai2026agentreadmesempiricalstudy}. Writing this guidance manually requires developers to identify relevant norms and express their applicability. Automating this work requires finding and interpreting evidence distributed across a repository~\cite{treude2024codecontributionprocessesinformation}. Explicit sources include contribution guides~\cite{prana2018categorizingcontentgithubreadme}, configurations, and workflows~\cite{elazhary2019idoisay}; source-code practices supply implicit evidence~\cite{allamanis2014miningidiomssourcecode}. A norm may have both explicit and implicit support, with source code and tests qualifying the applicability of a documented rule. The evidence's location need not define the norm's scope: a helper can constrain callers elsewhere. A recurring pattern may instead be confined to one component. Automated acquisition must determine which observations support a candidate norm, what conditions or exceptions apply, and which changes it governs before delivering the guidance to an agent.

Existing automated retrieval and documentation approaches primarily provide repository context~\cite{shrivastava2023repofusiontrainingcodemodels,hoang-etal-2026-codewiki}. Repoformer selectively retrieves context for repository-level code completion~\cite{pmlr-v235-wu24a}. Documentation systems such as DeepWiki and CodeWiki explain repository structure and behavior~\cite{hoang-etal-2026-codewiki}. Such context can help agents implement a task but may leave unclear which aspects contributors are expected to preserve and which additional artifacts they must deliver. A study of developer-written and automatically generated instruction files finds that agents generally follow their instructions without a consistent improvement in task resolution across the evaluated settings~\cite{gloaguen2026agentsmd}. This observation motivates assessing the quality and usefulness of guidance separately from whether agents follow it. These resources can contain norms, but acquiring them as reliable guidance still requires checking their content and applicability against repository evidence.

\textbf{We present \app, a task-independent approach to \textit{repository norm acquisition} for coding agents.} Before a concrete task begins, \app extracts candidates from explicit sources and infers candidates from source-code practices. It checks candidates against repository evidence, examining support and counterexamples to verify the proposed norm together with its scope, conditions, and exceptions. Bounded queries to local Git history supply additional evidence when needed to resolve questions about existing candidates. It then delivers selected norms and their supporting evidence through a norm package.

Preparing norms independently of a coding task allows the package to support multiple tasks and coding models for the repository. The static package organizes norms by applicability path, so agents can use native file tools to find guidance for the files and directories involved in a change. Each norm retains its applicability and evidence for inspection. The consuming agent selects relevant guidance and remains responsible for code generation, editing, and test execution.

To evaluate repository norm compliance and functional correctness, we construct RepoNormBench from CooperBench~\cite{khatua2026cooperbenchcodingagentsteammates}. It contains 121 coding tasks, with refined task descriptions, strengthened functional tests, and repository-grounded reference norms paired with automated compliance checks. Norm checks assess adherence to applicable repository norms, while functional tests assess the requested behavior. This distinction captures changes that solve a task but leave contribution obligations unmet~\cite{yu2026doespassratetell}. We compare \app guidance with no additional generated guidance (Raw) and with CodeWiki documentation, and assess norm quality through precision and coverage of the reference norms.

Across DeepSeek, Qwen, and Gemini, \app increases Overall Norm Compliance Rate (NCR) by 7.42--10.77\%, Contribution NCR by 31.64--45.44\%, and Prompt-omitted NCR by 11.34--17.69\%, all relative improvements over Raw. It improves these three measures over CodeWiki documentation for every model, with entirely positive paired 95\% confidence intervals. The larger gains concern contribution requirements less directly tied to functional correctness; changes in Behavioral NCR are smaller. Observed functional and joint success rates are 5.79--9.09 and 0.83--4.96 percentage points higher than with Raw, respectively, although paired comparisons do not reach the significance threshold after Holm correction. CodeWiki's functional results alongside its less consistent compliance gains suggest that context supporting task completion and explicit guidance for contribution requirements provide distinct benefits. Under its default configuration, \app achieves 86.00\% sampled precision in norm acquisition.

The paper contributes:
\begin{itemize}[beginpenalty=10000]
\item \app, a repository norm acquisition method and implementation combining explicit extraction, implicit inference, and verification of candidate content and applicability, with evidence retained in agent-readable norm packages.
\item RepoNormBench, a benchmark derived from CooperBench with refined task descriptions, strengthened functional tests, and repository-grounded norm-compliance metrics that complement functional evaluation.
\item An evaluation across three coding models that measures repository norm compliance, including adherence to norms not explicitly stated in task descriptions, alongside functional correctness and norm quality.
\end{itemize}

\section{Motivation}
\label{sec:background}

\begin{figure}[htbp]
    \centering
    \includegraphics[width=\linewidth]{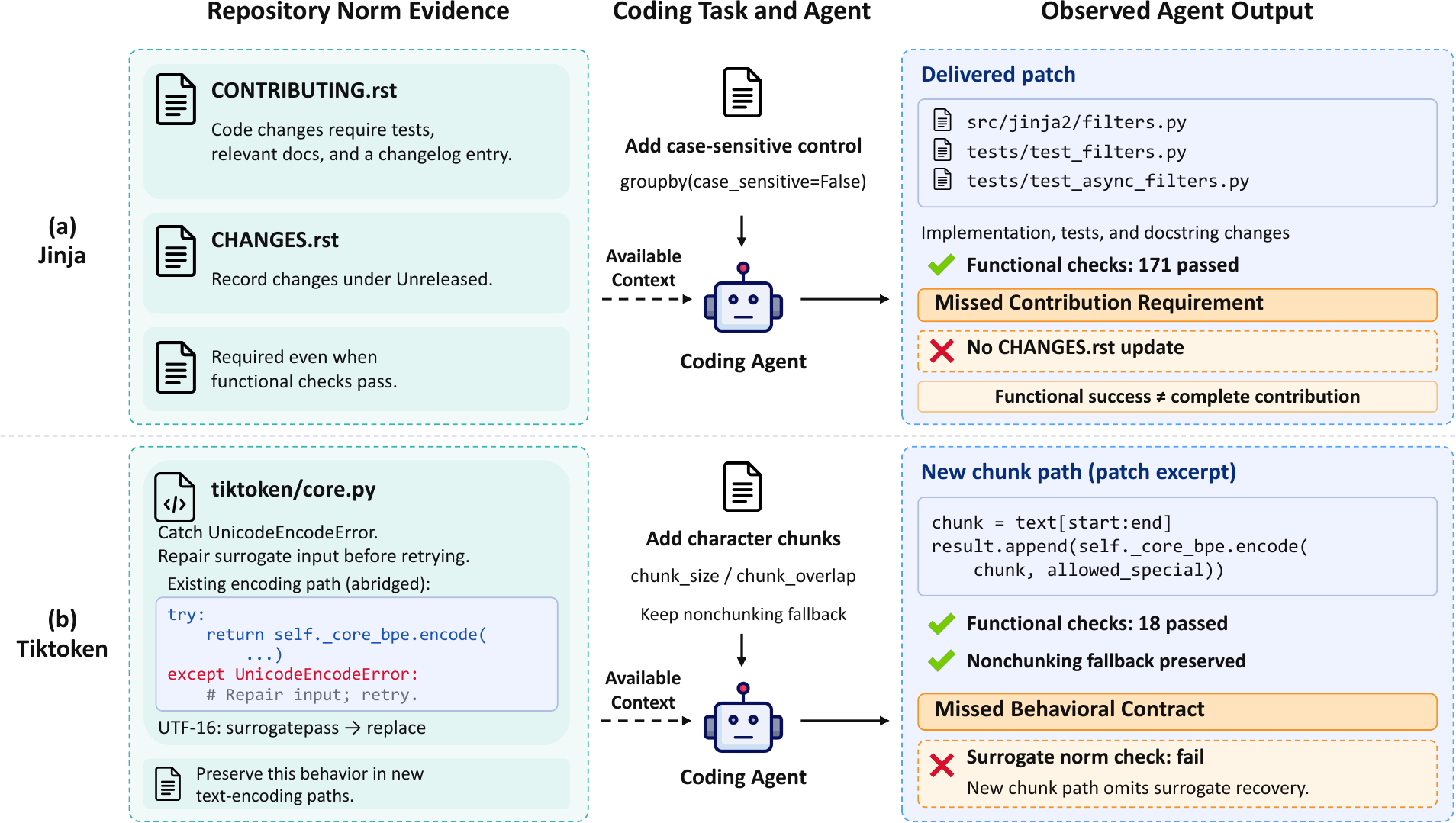}
    \caption{Agent outputs on two tasks. (a) Jinja passes 171 functional checks but omits a required changelog entry. (b) Tiktoken passes 18 functional checks but omits surrogate recovery in the new chunking path.}
    \Description{Two rows show repository norm evidence, a coding task, and observed agent output. In Tiktoken, existing source code catches UnicodeEncodeError, repairs surrogate-containing input, and retries encoding. The task adds character chunks while preserving the nonchunking fallback. The observed patch preserves that fallback but directly calls the Rust encoder in the new chunk path without recovery. It passes 18 functional checks and fails the surrogate norm check. In Jinja, CONTRIBUTING.rst requires tests, relevant documentation updates, and a changelog entry. The observed patch adds case-sensitive control to groupby and changes the implementation, synchronous and asynchronous tests, and docstrings. It passes 171 functional checks but does not update CHANGES.rst.}
    \label{fig:motivation}
\end{figure}

Acquiring repository norms poses two challenges: discovering relevant norms beyond the task description and establishing where they apply. Figure~\ref{fig:motivation} illustrates these challenges with observed agent outputs from two RepoNormBench tasks in the Raw condition, without additional generated repository guidance. Both patches passed the functional checks for their tasks, yet the Jinja patch omitted a contribution requirement and the Tiktoken patch omitted a behavioral contract.

\subsection{Challenge 1: Discovering Norms Beyond the Task Description}
\label{sec:motivation-discovery}

Jinja's \texttt{CONTRIBUTING.rst} requires code changes to include tests, updates to relevant documentation and docstrings, and an entry in \texttt{CHANGES.rst}. It also requires inline \texttt{versionchanged} annotations in relevant docstrings. The changelog's current Unreleased section provides the destination for new entries. These contribution requirements apply when an agent changes a user-facing filter, even if the task description specifies only its runtime behavior.

For task \texttt{JIN-F01}, Qwen 3.5 Flash extended \texttt{groupby} with \texttt{case\_sensitive=False} in synchronous and asynchronous environments. Its patch modified \texttt{src/jinja2/filters.py}, added tests in \texttt{tests/test\_filters.py} and \texttt{tests/test\_async\_filters.py}, and updated the docstring, including a \texttt{versionchanged} marker. All 171 functional checks for the task passed, but the patch did not update \texttt{CHANGES.rst}, as shown in Figure~\ref{fig:motivation}(a). The changelog requirement was not stated in the task description.

The acquisition challenge is to locate contribution requirements among repository sources and identify the artifacts each requirement calls for. In this case, the contribution guide states the obligation, while the changelog shows where to fulfill it. Providing tests and docstring changes does not satisfy the separate changelog obligation. Acquisition must distinguish requirements from descriptive material and preserve obligations that functional checks do not assess.

\subsection{Challenge 2: Establishing Applicability}
\label{sec:motivation-applicability}

Tiktoken uses a Rust BPE implementation to encode text into tokens. In \texttt{tiktoken/core.py}, \texttt{Encoding.encode} catches \texttt{UnicodeEncodeError} when the Rust encoder cannot accept surrogate-containing input. It repairs the input by encoding it as UTF-16 with \texttt{surrogatepass} and decoding it with \texttt{replace}, then retries. A source comment explains why this recovery handles surrogate pairs and lone surrogates. New text-encoding paths that call the same Rust encoder need to preserve this behavior.

For task \texttt{TIK-F04}, DeepSeek V4 Flash added character-window encoding through \texttt{chunk\_size} and \texttt{chunk\_overlap}. The task explicitly required preserving the ordinary Unicode fallback on the nonchunking path, which the patch retained. The new chunking branch directly called the Rust encoder without the recovery, as shown in Figure~\ref{fig:motivation}(b). The patch passed all 18 functional checks for the task but failed the surrogate-recovery norm check. The omission occurred in the added path, where the existing encoding contract also applied.

The acquisition challenge is to turn an implementation fact into guidance with justified conditions and scope. Here, surrogate-containing text at the Rust encoding boundary determines applicability; the evidence in the existing path also supports guidance for new paths crossing that boundary. More generally, verification must examine whether a candidate's proposed norm and qualifiers are supported, considering counterexamples, allowed exceptions, and uses outside its intended scope. When a material condition remains unresolved, acquisition must reflect that uncertainty in its judgment and delivery.

\app addresses these challenges through explicit extraction and implicit inference, followed by unified verification and norm packaging.

\section{Approach}
\label{sec:approach}
\begin{figure}[htbp]
    \centering
    \includegraphics[width=0.95\columnwidth]{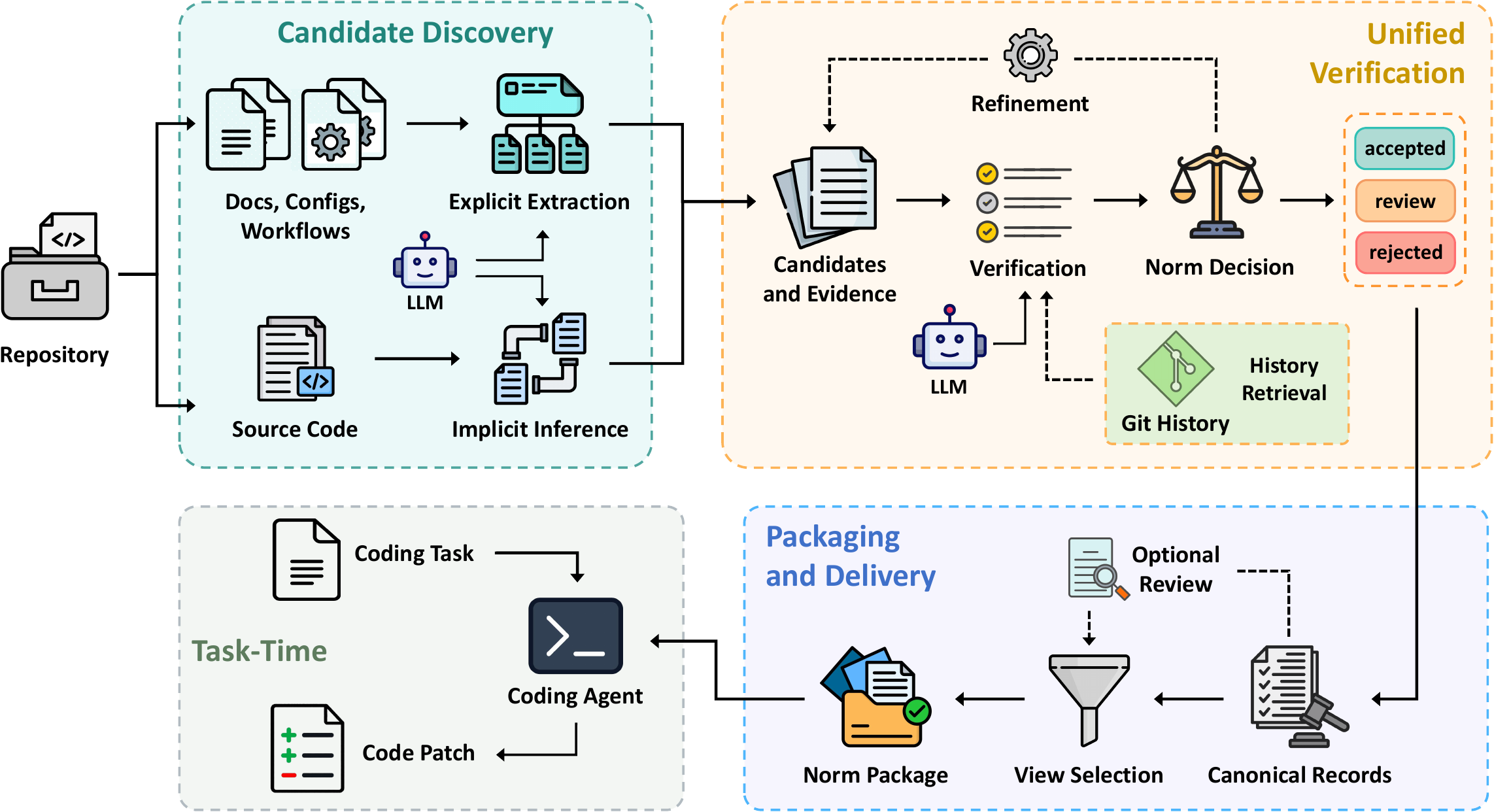}
    \Description{A repository feeds explicit norm extraction and implicit norm inference. Candidates and evidence enter unified verification, with Git retrieval when needed. The resulting repository norm package can undergo optional review. During coding task execution, an existing coding agent consumes the norm package and a coding task to produce a code patch.}
    \caption{Overview of \app.}
    \label{fig:workflow}
\end{figure}

\app acquires repository-specific norms before a coding task begins. Figure~\ref{fig:workflow} shows its three main stages: discovering norm candidates through explicit extraction and implicit inference, verifying their content and applicability against repository evidence, and delivering selected norms in a static package. An existing coding agent chooses which norms to consult while carrying out a task.

\subsection{Overview and Norm Representation}
\label{sec:approach-overview}

The input is a repository $R$. The explicit channel extracts what repository sources state or configure; the implicit channel proposes conventions from source observations and checks them against an explicitly defined code population. Both produce candidates with evidence and a channel-local assessment of whether the candidate and its applicability are supported. Unified verification determines their final status.

We represent the semantics of a norm as
\begin{equation}
    n = (r,s,c,e),
    \label{eq:norm-semantics}
\end{equation}
where $r$ states the governed behavior, $s$ identifies the governed artifacts using repository-relative paths or glob patterns, $c$ specifies an activation condition, and $e$ records an exception, if any. Evidence provenance separately identifies where support was found: a test may provide evidence about a production API without restricting the norm to test files.

Unified verification produces canonical norm records that associate these semantics with supporting evidence, counterevidence, and a decision status: \texttt{accepted}, \texttt{review}, or \texttt{rejected}. The assigned status is a system judgment whose correctness is evaluated separately. Each record also specifies the norm's authority, termed normative strength, and whether and how it is supplied to an agent, governed by a delivery policy (Section~\ref{sec:approach-delivery}).

\subsection{Explicit Norm Extraction}
\label{sec:approach-explicit}

\begin{figure}[htbp]
    \centering
    \includegraphics[width=0.85\linewidth]{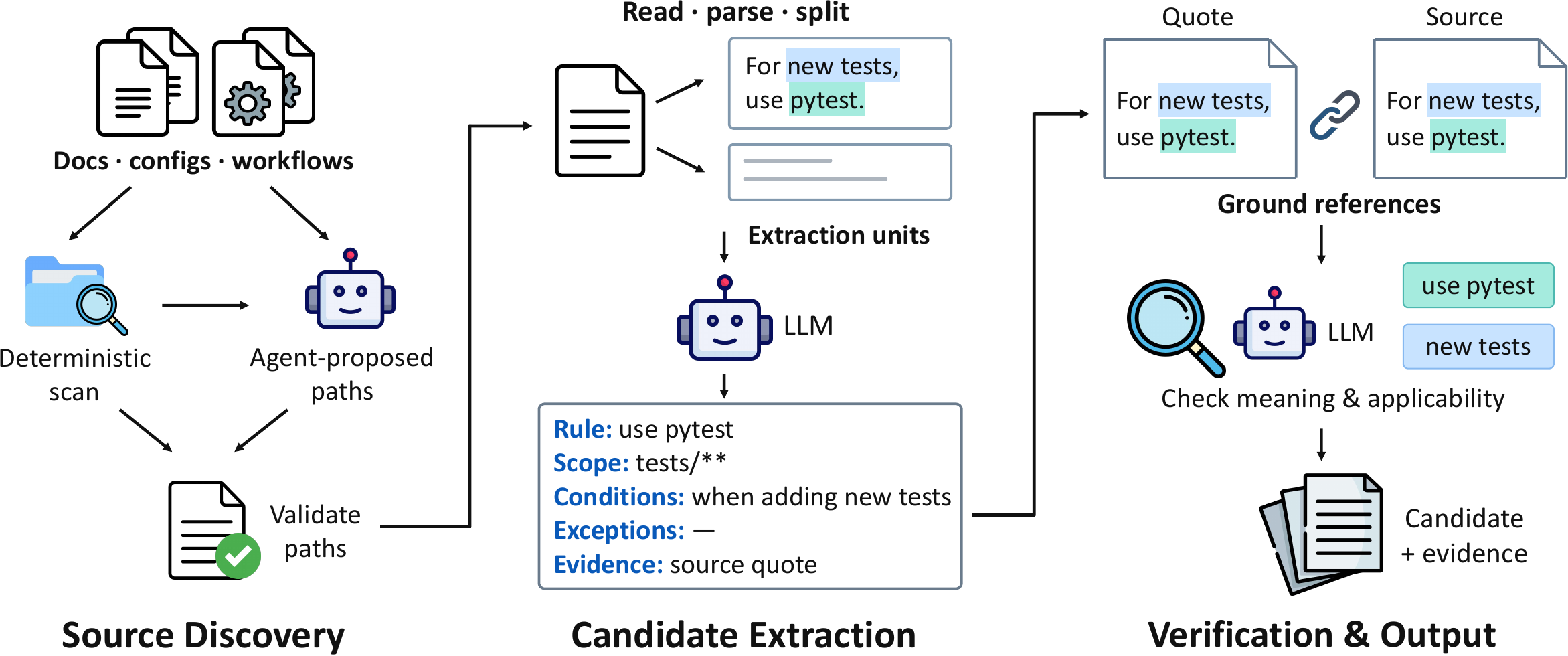}
    \caption{Explicit norm extraction through source discovery, candidate extraction, and verification.}
    \Description{Three stages show source discovery, candidate extraction, and verification with output. A deterministic scan and agent-proposed paths locate documents, configurations, and workflows. Validated sources are read, parsed, and split into units. A language model extracts a rule, scope, conditions, exceptions, and evidence. The candidate requires pytest, has scope tests/** and the condition when adding new tests, and has no exception. Source grounding checks the quote against the original content, and a semantic check examines the rule and its applicability before passing the candidate and evidence to unified verification.}
    \label{fig:explicit-extraction}
\end{figure}

The explicit channel turns repository declarations into candidates while preserving their intended applicability. Figure~\ref{fig:explicit-extraction} illustrates the process with the statement ``For new tests, use pytest.''

\paragraph{\textbf{Source discovery.}}
A deterministic scan locates repository instructions, documentation, configurations, and workflows. A bounded discovery agent proposes additional repository paths. A host program coordinates \app's deterministic operations and model calls. It validates the proposed paths before admitting their content for extraction.

\paragraph{\textbf{Candidate extraction.}}
\app reads, parses, and splits the sources into extraction units. A language model extracts each candidate's rule, scope, conditions, exceptions, and evidence references. The example in Figure~\ref{fig:explicit-extraction} assumes tests are stored under \texttt{tests/}. Its candidate records ``use pytest,'' the scope \texttt{tests/**}, and the condition ``when adding new tests,'' with no exception.

\paragraph{\textbf{Verification and output.}}
\app first grounds evidence references in the original text, parsed configuration values, or workflow commands. A separate model check assesses whether the evidence supports the rule, path scope, and activation condition. A candidate requiring pytest for all existing tests would exceed that statement even if its quote were correctly grounded. This check also distinguishes requirements, recommendations, and descriptive facts; a single workflow occurrence does not establish a repository-wide obligation. The channel passes candidates, grounded evidence, and local assessments to unified verification. Discovery gaps and candidate verification failures are recorded separately.

\subsection{Implicit Norm Inference}
\label{sec:approach-implicit}

\begin{figure}[htbp]
    \centering
    \includegraphics[width=0.95\linewidth]{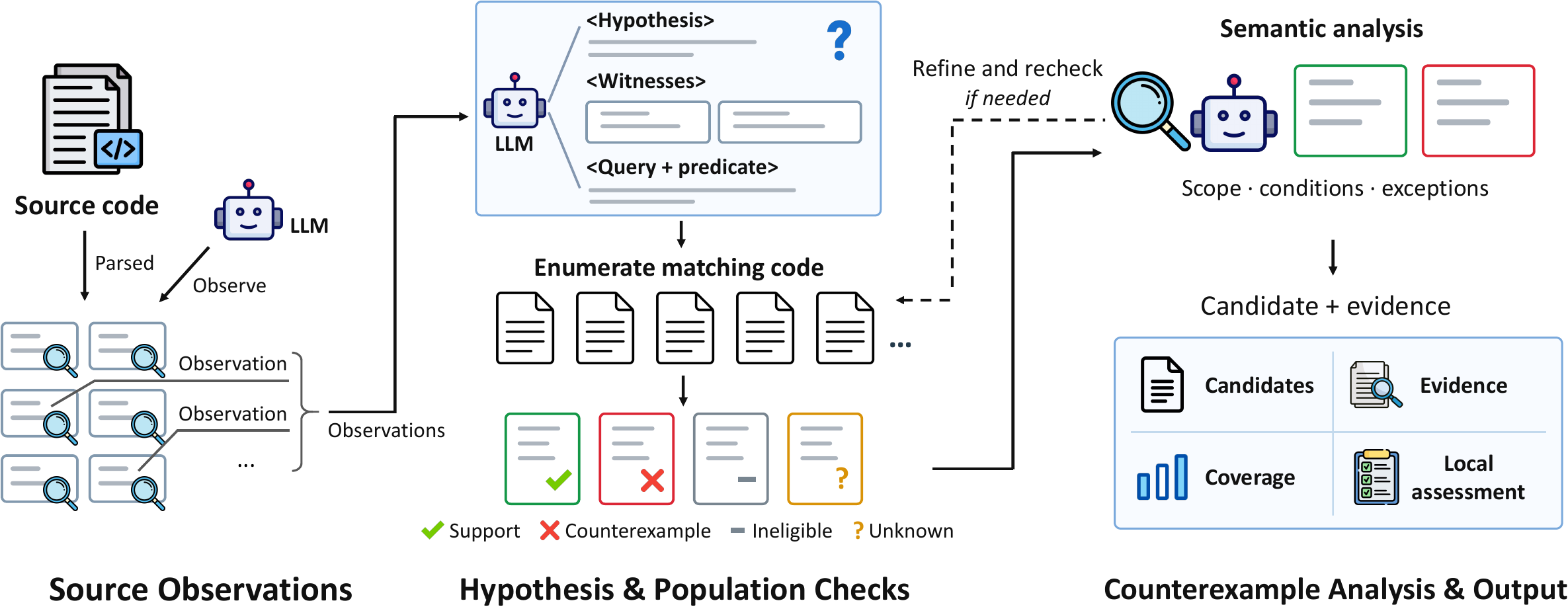}
    \caption{Implicit norm inference through source observations, population checks, and counterexample analysis, with at most one local refinement.}
    \Description{Three stages show source observations, hypothesis and population checks, and counterexample analysis with output. Structural and semantic observations motivate a hypothesis with witnesses, a population query, and a checking predicate. Matching code is enumerated, and inspected members are labeled support, counterexample, ineligible, or unknown. Semantic analysis examines scope, conditions, and exceptions. A dashed refinement arrow returns to population checks if needed. The output contains candidates, evidence, coverage, and local assessments.}
    \label{fig:implicit-inference}
\end{figure}

The implicit channel infers conventions from current source code by testing hypotheses against an explicitly defined code population. Figure~\ref{fig:implicit-inference} summarizes the inference workflow. We explain its stages below using Jinja's empty-input handling as a running example: several filters return a configured undefined value when no input element is available, but this behavior does not extend to every filter that consumes a sequence.

\paragraph{\textbf{Source observations.}}
The Source Observations stage on the left of Figure~\ref{fig:implicit-inference} combines structural indexing with semantic observation extraction. \app parses source code into abstract syntax trees and indexes files, code entities, call and import sites, and relations, retaining source locations. It also organizes code into bounded units, such as functions and classes with relevant context, for a language model to interpret. In Jinja, the \texttt{first} filter catches \texttt{StopIteration}, while \texttt{random} catches \texttt{IndexError}; both return a value through the environment's configured \texttt{undefined} factory. An undefined value represents a missing result, with behavior controlled by the environment. Their semantic observations capture the shared empty-input behavior despite the different selection operations. Each observation remains linked to its supporting code.

\paragraph{\textbf{Hypotheses and population checks.}}
A separate model stage examines bounded sets of observations to propose norm hypotheses. The Jinja observations suggest an initial hypothesis: filters that reduce a sequence to a single value return the configured undefined value on empty input. In this example, the \texttt{first} and \texttt{random} observations serve as hypothesis witnesses, corresponding to the ``Witnesses'' entry in Figure~\ref{fig:implicit-inference}. The ``Query + predicate'' entry specifies where and how to test the hypothesis. A query enumerates function entities in \path{src/jinja2/filters.py}, and a semantic predicate asks whether each function implements a sequence-to-value filter and whether its empty-input path returns the configured undefined value. The host validates and executes the query over the source index.

At ``Enumerate matching code'' in Figure~\ref{fig:implicit-inference}, the host enumerates query matches before selecting members for inspection. In the Jinja example, these are function entities, whose code is inspected for eligibility and support. Given a complete enumeration, let $P_n$ be the matched population, $I_n$ the inspection set, and $U_n$ the uninspected members:
\begin{equation}
    I_n \subseteq P_n, \qquad U_n=P_n\setminus I_n,
    \qquad |P_n|=|I_n|+|U_n|.
    \label{eq:population-coverage}
\end{equation}
A structural predicate determines eligibility and support through source-index queries, permitting a deterministic census when those queries are complete. The Jinja hypothesis requires semantic checks: the presence of an \texttt{undefined} call alone does not establish the empty-input behavior. The host selects a census or a deterministic stratified sample within the inspection budget. Each inspected member receives one of the four labels below the matching code in Figure~\ref{fig:implicit-inference}. The \texttt{first}, \texttt{last}, \texttt{random}, \texttt{min}, and \texttt{max} implementations support the hypothesis; \texttt{sum} and \texttt{join} provide counterexamples. Unrelated functions are ineligible, and unresolved cases are labeled unknown.

\paragraph{\textbf{Counterexample analysis and output.}}
When inspected members provide support and required technical checks are complete, a model examines representative support, counterexamples, and uncertain members. The ``Semantic analysis'' step on the right of Figure~\ref{fig:implicit-inference} checks repository specificity, contributor value, and applicability. In Jinja, \texttt{sum} returns its initial value, normally zero, for an empty sequence, and \texttt{join} returns an empty string. These counterexamples show that the initial hypothesis is too broad: it includes filters with different empty-input semantics. Clone-related copies and shared helpers are accounted for when selecting representatives and assessing independent support. For example, \texttt{min} and \texttt{max} share a helper, so their matching behavior is not independent support from two separate implementations.

The dashed ``Refine and recheck'' arrow in Figure~\ref{fig:implicit-inference} permits one local refinement. For Jinja, the scope narrows to filters that select an element from their input: \texttt{first}, \texttt{last}, \texttt{random}, \texttt{min}, and \texttt{max}. The revised hypothesis returns to population construction and checking, retaining previously inspected observations; \texttt{sum} and \texttt{join} become ineligible under the narrower scope. The resulting candidate states that these selectors return the environment's configured undefined value for empty input within this scope. The ``Candidate + evidence'' output carries this rule, its applicability, inspected support, coverage, local assessment, and unresolved issues to unified verification.

\subsection{Unified Norm Verification}
\label{sec:approach-verification}

Unified verification gives candidates from both channels a final status under a common evidence policy. It assembles current evidence, obtains historical context when needed, and checks whether each candidate is supported as written. Algorithm~\ref{alg:norm-verification} summarizes the control flow.

\begin{algorithm}[t]
\caption{Candidate Verification and Decision}
\label{alg:norm-verification}
\resizebox{0.9\linewidth}{!}{%
\begin{minipage}{\linewidth}
\begin{algorithmic}[1]
\REQUIRE Repository $R$, admitted candidates $C$, verification policy
\ENSURE Canonical records with evidence, statuses, and revision links
\STATE $K \gets \operatorname{MakeCases}(C)$; $L \gets \emptyset$
\FOR{$t \in \{0,1\}$}
    \STATE $Q \gets \emptyset$ \COMMENT{Pending refinement proposals}
    \FORALL{$k \in K$}
        \STATE $E \gets \operatorname{AssembleCurrentEvidence}(R,k)$; $T \gets \emptyset$
        \IF{$\operatorname{HostResolvable}(k,E)$}
            \STATE $f \gets \operatorname{HostFinding}(k,E)$
        \ELSE
            \STATE $h \gets \operatorname{ValidateHistoryNeed}(k,E)$
            \STATE $T \gets \operatorname{RetrieveAndClassifyIfNeeded}(R,k,E,h)$
            \STATE $f \gets \operatorname{VerifyCase}(k,E,T)$
        \ENDIF
        \STATE $d \gets \operatorname{ApplyDecisionPolicy}(k,E,T,f)$
        \STATE $L \gets L \cup \{(k,E,T,f,d)\}$
        \IF{$t=0$}
            \STATE $Q \gets Q \cup \operatorname{ValidRefinements}(k,f)$
        \ENDIF
    \ENDFOR
    \IF{$t=0$}
        \STATE $K \gets \operatorname{ReacquireForChildren}(R,Q,L)$
    \ENDIF
\ENDFOR
\RETURN $\operatorname{CanonicalizeWithLineage}(L)$
\end{algorithmic}
\end{minipage}%
}
\end{algorithm}

\paragraph{\textbf{Current evidence and deterministic routing.}}
Each candidate forms a verification case containing its semantics and channel-local assessment. The host assembles grounded explicit evidence, inspected source members, counterexample findings, and bounded repository context. For the Jinja candidate in the example above, the bundle links the empty-input return paths to the selector scope and records inspection coverage. Evidence origins and dependencies preserve the distinction between separate implementations and shared helpers. This assembly corresponds to \texttt{AssembleCurrentEvidence} in Algorithm~\ref{alg:norm-verification}.

The host resolves cases with sufficient deterministic grounds before requesting another model judgment. The pytest candidate in Figure~\ref{fig:explicit-extraction} can take this path when its quote, path scope, and new-test condition pass the local checks, required evidence is complete, and no counterevidence, conflict, or historical inquiry remains. Missing required current evidence leads to \texttt{review}; admissible current counterevidence can support rejection under the same decision policy.

\paragraph{\textbf{Candidate-conditioned historical evidence.}}
A historical inquiry identifies a specific question about a candidate and binds it to current evidence and repository paths. For the Jinja candidate, the inquiry can ask whether the empty-input fallback was preserved when asynchronous filters were moved alongside their synchronous implementations. The host anchors retrieval to the filter's code, symbols, and \path{src/jinja2/filters.py}, then searches locally available Git ancestors within a bounded budget. Jinja's asynchronous-filter refactoring provides a relevant change: it renames the synchronous implementation to \texttt{sync\_do\_first} and moves an asynchronous \texttt{do\_first} into the same module, preserving its \texttt{StopAsyncIteration} handler and configured undefined return. This change supports preserving the fallback across the two implementations. A model classifies retrieved changes as support, contradiction, qualification, retirement or replacement, context, or uncertainty. Shared origins prevent the moved implementation and its introducing patch at the new location from counting as independent support.

The History Retrieval branch in Figure~\ref{fig:workflow} supplies this evidence to verification. The historical Jinja patch supplements the current synchronous and asynchronous return paths; missing supplementary history need not block acceptance. Adverse historical findings prompt rechecking against current evidence. History can also help resolve a material ambiguity. If a historical question is required for the decision and remains unresolved, the candidate remains in \texttt{review}. A bounded search with no match is recorded separately from unavailable or truncated history.

\paragraph{\textbf{Semantic judgment and decision policy.}}
For cases requiring semantic judgment, a model acts as the case verifier. It assesses the rule and its applicability using the evidence bundles, coverage, and individual channel judgments, such as support for the rule and its scope. The channel's overall conclusion is withheld to avoid anchoring the verifier to that earlier judgment. The verifier cites supplied evidence, and the host applies a deterministic policy to its findings.

Applicable integrity, dependency, and conflict checks must pass before acceptance or rejection. Acceptance requires grounded current support, complete required checks, and resolution of material conflicts and local uncertainties. The narrowed Jinja candidate is \texttt{accepted} once the selector scope and empty-input semantics are supported and these checks pass; an unverified asynchronous path claimed by the rule leaves it in \texttt{review}. Rejection requires a failed core rule-support judgment backed by candidate-bound current counterevidence: a rule specifically requiring \texttt{sum} to return an undefined value is refuted by its return of the initial value. Scope or qualifier mismatches and unresolved interpretations or disagreements lead to \texttt{review}. Historical evidence cannot independently establish acceptance or rejection.

\paragraph{\textbf{Bounded candidate refinement.}}
Unified verification permits one refinement round, separate from the implicit channel's local refinement. A revision may clarify wording, scope, conditions, or exceptions while preserving the core action and direction. For example, Jinja's \texttt{last} requires a reversible input. If an incoming candidate extends its fallback to all iterables, the verifier can add this condition while preserving the undefined-return rule. The revision requires renewed evidence collection and verification, including checks of newly included code if applicability expands. Previously inspected observations are retained. The \texttt{ReacquireForChildren} step in Algorithm~\ref{alg:norm-verification} binds the revised candidate to fresh checks and its parent; further revision requests or incomplete rechecking lead to \texttt{review}.

\subsection{Norm Packaging and Delivery}
\label{sec:approach-delivery}

\app selects entries from the canonical records to form an agent-facing norm view: the norms made available to a coding agent. The default view contains accepted records and is called the Accepted-Only View.

Normative strength distinguishes an applicable requirement or enforced configuration (required), an explicit recommendation (recommended), and a supported source convention (observed). The Jinja empty-input norm has observed strength because its support comes from source behavior. Delivery policy accounts for applicability conditions and execution risk. Conditional entries are intended for use only when their stated conditions hold; reference-only entries remain readable without being offered as default action guidance.

\paragraph{\textbf{Package structure.}}
The selected view is delivered as a static Markdown package. A navigation entry (\texttt{START.md}) links to norm pages organized by repository path and to a complete catalog. A detail page for the Jinja norm retains the selector scope, empty-input condition, and source locations of the undefined-return paths. Each page preserves the norm's full semantics and supporting evidence. The catalog offers an alternative browsing route and includes reference-only entries.

\paragraph{\textbf{Organization by applicability path.}}
\app groups norms selected for action guidance by the repository paths in their applicability scopes. The Jinja norm is indexed at \path{src/jinja2/filters.py}, where its governed filters are implemented. Wildcard scopes use the directory prefix preceding the first wildcard, so \texttt{tests/**} is indexed under \texttt{tests/}; patterns without a fixed prefix are indexed at the repository root. Norms with multiple scope entries can appear at several locations, all linking to the same detail record. This placement follows applicability, with evidence-source paths retained separately as provenance. Path pages preserve the complete rule, scope, conditions, exceptions, and normative strength.

\paragraph{\textbf{Consumption.}}
Before a task begins, the host checks that the selected view matches the target repository and supplies it as read-only files. After reading the entry document, a coding agent changing a Jinja selector can follow the path page for \path{src/jinja2/filters.py} to the empty-input norm and inspect its supporting code. The agent may also search the package or consult the catalog. It selects the norms to apply and performs the task's generation and execution loop. Internal audit records remain outside the delivered view.

\paragraph{\textbf{Optional developer review.}}
A web interface lets developers and repository maintainers inspect norms with \texttt{review} status and their evidence, accept or reject individual items, and publish their selections. Publishing a session produces an Accepted-after-Review View, which combines original accepted records with review items explicitly accepted in that session, subject to delivery policy. Untouched items remain unprocessed, and session decisions do not rewrite canonical statuses.
The developer-review interface is not used in the downstream coding experiments; those experiments use the Accepted-Only View. The Accepted-after-Review View evaluated for norm quality is instead produced through a separate automated review process.

\section{Evaluation}
\label{sec:evaluation}
We evaluate whether repository norms help coding agents follow repository practices, including contribution requirements that functional tests do not directly assess. Four research questions address norm compliance as the primary outcome, functional correctness as a secondary outcome, norm quality, and the time and token costs of norm acquisition and guided coding. Detailed evaluation data, including results not tabulated here, are available in our public repository\footnote{\url{https://github.com/SYSUSELab/RepoNorm}}.

\subsection{Experimental Setup}
\label{sec:eval-setup}

\paragraph{\textbf{RepoNormBench.}}
To keep computational costs and repository-grounded norm annotation manageable, we select six of CooperBench's nine Python repositories, prioritizing smaller codebases~\cite{khatua2026cooperbenchcodingagentsteammates}. RepoNormBench comprises 121 coding tasks across 14 versions of these repositories (Table~\ref{tab:eval-setup}). Each task asks one agent to implement one feature and supplies a feature description and a repository checkout. An evaluation-owned test suite determines functional success.

We refine the descriptions of all 121 tasks to reduce ambiguous wording and remove some implementation guidance. We also strengthen the functional tests, adding 501 test-function identities, removing 503, and modifying 78 retained functions. With assistance from GPT-5.6-Sol and GPT-6-Astra, we add norm-compliance metrics based on repository-grounded reference norms and automated checks to evaluate repository practices separately from functional correctness. Within each coding model, all three conditions use the same final task descriptions and functional tests. Throughout this process, we do not consult the norm packages acquired by \app.

The norm reference contains 160 distinct norms with 1,024 task--norm associations. Model-assisted source review grounds norms in repository documentation, configuration, source code, and tests, and records their conditions and task applicability. We classify each norm into one of 11 categories: authored tests, documentation, changelogs, inline changes, style, typing, compatibility, dependencies, public exports, public interfaces, and runtime behavior. A norm can apply to multiple tasks. For each applicable task--norm pair, we also annotate whether the task description explicitly states the norm. Norm checks are calibrated against compliant reference implementations and violating variants; Section~\ref{sec:eval-compliance} describes the automated scoring and reliability analyses. The reference covers norms relevant to these tasks. Hidden tests, reference solutions, and reference norms are reserved for evaluation and excluded from norm acquisition and the agents' task inputs.

% Generated by artifacts/evaluation/build_tables.py from final exports.
\newsavebox{\evalSetupTable}
\begin{table}[htbp]
\centering
\caption{RepoNormBench repositories, tasks, and reference norms.}
\label{tab:eval-setup}
\small
\sbox{\evalSetupTable}{%
\resizebox{0.95\width}{!}{%
\begin{tabular}{@{}lcccc@{}}
\toprule
Repository & Tasks & Versions & \makecell{Reference norms} & \makecell{Total text lines (k)} \\
\midrule
Outlines & 22 & 3 & 23 & 24.83--41.28 \\
DSPy & 23 & 3 & 24 & 51.75--57.58 \\
Tiktoken & 10 & 1 & 6 & 3.00 \\
Click & 27 & 3 & 38 & 25.64--27.81 \\
Jinja & 30 & 3 & 59 & 28.26--28.31 \\
Dirty Equals & 9 & 1 & 10 & 4.53 \\
\midrule
\textbf{Overall} & \textbf{121} & \textbf{14} & \textbf{160} & -- \\
\bottomrule
\end{tabular}%
}%
}
\usebox{\evalSetupTable}
\par\smallskip
\begin{minipage}{\wd\evalSetupTable}
\footnotesize
\raggedright
Versions denote distinct base commits. Text lines include code and non-code text; ranges span versions ($k=1{,}000$).
\end{minipage}
\end{table}

\paragraph{\textbf{Compared conditions.}}
We compare three conditions for each coding model. \textit{Raw} supplies the task and repository without an additional generated guidance package. \textit{CodeWiki} supplies generated repository documentation. Its published evaluation on CodeWikiBench reports higher average documentation quality than DeepWiki, deepwiki-open, and OpenDeepWiki~\cite{hoang-etal-2026-codewiki}. We select CodeWiki because, like \app, it generates a task-independent repository-level resource that can be reused across coding tasks. \textit{\app} supplies the Accepted-Only View described in Section~\ref{sec:approach-delivery}. Both generated resources use DeepSeek V4.1 Flash with thinking disabled and without access to task descriptions, reference solutions, or hidden tests. Tasks reuse the resources prepared for their repositories. CodeWiki exposes its documentation directory; \app exposes a Markdown entry point and norm files. The coding agent chooses which material to read.

\paragraph{\textbf{Models and execution.}}
We use mini-swe-agent v2.1.0~\cite{yang2024sweagent} as the coding harness and evaluate DeepSeek V4 Flash (0423), Qwen 3.5 Flash, and Gemini 3.1 Flash-Lite. DeepSeek and Gemini use high and medium reasoning effort, respectively. Qwen uses thinking mode with a budget of 8,192 reasoning tokens per model call. All three models use temperature zero and a limit of 300 model calls per task. Each model is evaluated on all 121 tasks under all three conditions.

\subsection{RQ1: Norm Compliance}
\label{sec:eval-compliance}

\textit{To what extent do coding-agent solutions comply with repository-specific norms?}

\paragraph{\textbf{Design.}}
We apply automated norm checks to the solutions produced for all 121 tasks under each model and condition. We evaluate these same solutions for functional correctness in RQ2 (Section~\ref{sec:eval-correctness}). These checks evaluate observable behavior, static properties, and patch contents against applicable reference norms. Runtime checks test required behavior and public exports; static tools check formatting or typing requirements. For norms requiring authored tests, a passing decision requires identifiable new or modified tests delivered with the patch and a successful test run on the patched repository. Documentation checks identify the required document changes.

\textbf{NCR measures.} For task $i$ and analysis scope $s$, let $P_{i,s}$ and $F_{i,s}$ count applicable norms receiving passing and failing decisions. With $\mathcal{T}_s$ denoting tasks with at least one decided norm in that scope, Norm Compliance Rate (NCR) is
\begin{equation}
    \mathrm{NCR}_s=\frac{1}{|\mathcal{T}_s|}\sum_{i\in\mathcal{T}_s}\frac{P_{i,s}}{P_{i,s}+F_{i,s}}.
    \label{eq:norm-compliance}
\end{equation}
Each task has equal weight, separately for each model and condition. Inapplicable norms and unresolved decisions do not enter $P_{i,s}+F_{i,s}$; tasks with no decided norms are excluded from NCR.

We report four NCR measures. \textit{Overall NCR} summarizes all evaluated categories. \textit{Contribution NCR} covers six categories: authored tests, documentation, changelogs, inline changes, style, and typing. These contribution requirements are less directly tied to functional correctness: implementing the requested behavior need not satisfy the accompanying contribution obligations. \textit{Behavioral NCR} covers the remaining five categories: runtime behavior, public exports, public interfaces, compatibility, and dependencies. These requirements are more directly connected to functional behavior and are often addressed while implementing or preserving correctness. \textit{Prompt-omitted NCR} covers norms labeled as not explicitly stated in the task description, assessing compliance beyond explicit task instructions.

\textbf{Statistical comparisons.} For each model and NCR scope, we compare \app with Raw and CodeWiki using paired NCR differences. Only tasks with a nonzero decided denominator in both conditions enter a comparison. We resample task pairs 100,000 times and report two-sided bias-corrected and accelerated (BCa) 95\% bootstrap confidence intervals for the mean NCR difference. Each interval concerns one comparison; we do not adjust these intervals for simultaneous coverage. Because the paired task set can differ from each condition's descriptive population, the paired effect need not equal the difference between the means in Table~\ref{tab:eval-compliance}.

% Generated by artifacts/evaluation/build_statistics.py.
\newsavebox{\evalComplianceTable}
\begin{table}[htbp]
\centering
\caption{Norm compliance across four scopes.}
\label{tab:eval-compliance}
% \small
\sbox{\evalComplianceTable}{%
\resizebox{0.95\width}{!}{%
\begin{tabularx}{\linewidth}{@{}cl*{5}{>{\centering\arraybackslash}X}@{}}
\toprule
\multirow{2}{*}{Model} & \multirow{2}{*}{Scope} & Raw & \multicolumn{2}{c}{CodeWiki} & \multicolumn{2}{c}{\textbf{\app}} \\
\cmidrule(lr){3-3}\cmidrule(lr){4-5}\cmidrule(lr){6-7}
 &  & NCR (\%) & NCR (\%) & $\Delta$ (\%) & NCR (\%) & $\Delta$ (\%) \\
\midrule
\multirow{4}{*}{DeepSeek} & Overall & 65.58 & 65.66 & \evalup{0.12} & \textbf{72.64} & \evalup{10.77} \\
 & Contribution & 33.82 & 34.75 & \evalup{2.76} & \textbf{49.19} & \evalup{45.44} \\
 & Behavioral & \textbf{96.46} & 95.45 & \evaldown{1.04} & 96.03 & \evaldown{0.44} \\
 & Prompt-omitted & 52.07 & 52.41 & \evalup{0.64} & \textbf{61.28} & \evalup{17.69} \\
\midrule
\multirow{4}{*}{Qwen} & Overall & 63.93 & 62.18 & \evaldown{2.72} & \textbf{68.67} & \evalup{7.42} \\
 & Contribution & 32.15 & 29.31 & \evaldown{8.84} & \textbf{43.39} & \evalup{34.94} \\
 & Behavioral & \textbf{94.75} & 93.03 & \evaldown{1.81} & 93.77 & \evaldown{1.03} \\
 & Prompt-omitted & 50.12 & 48.53 & \evaldown{3.18} & \textbf{55.81} & \evalup{11.34} \\
\midrule
\multirow{4}{*}{Gemini} & Overall & 57.61 & 58.36 & \evalup{1.31} & \textbf{63.38} & \evalup{10.03} \\
 & Contribution & 25.28 & 25.63 & \evalup{1.36} & \textbf{33.28} & \evalup{31.64} \\
 & Behavioral & 87.60 & 87.84 & \evalup{0.27} & \textbf{90.28} & \evalup{3.05} \\
 & Prompt-omitted & 43.36 & 42.79 & \evaldown{1.30} & \textbf{49.68} & \evalup{14.59} \\
\bottomrule
\end{tabularx}%
}%
}
\usebox{\evalComplianceTable}
\par\smallskip
\begin{minipage}{\wd\evalComplianceTable}
\footnotesize
\raggedright
Cells include 119--121 tasks with decided norms. $\Delta$: relative change from Raw. Bold: highest NCR per model and scope.
\end{minipage}
\end{table}

\paragraph{\textbf{Results.}}
Table~\ref{tab:eval-compliance} summarizes compliance across the four NCR measures.

\textbf{Overall compliance.} \app attains the highest Overall NCR for all three models, with relative improvements of 7.42--10.77\% over Raw and 8.61--10.64\% over CodeWiki. CodeWiki's Overall NCR changes by $+0.12\%$, $-2.72\%$, and $+1.31\%$ relative to Raw for DeepSeek, Qwen, and Gemini, respectively. Its observed functional gains (Section~\ref{sec:eval-correctness}) do not consistently carry over to norm compliance.

\begin{figure}[htbp]
    \centering
    \includegraphics[width=0.75\linewidth]{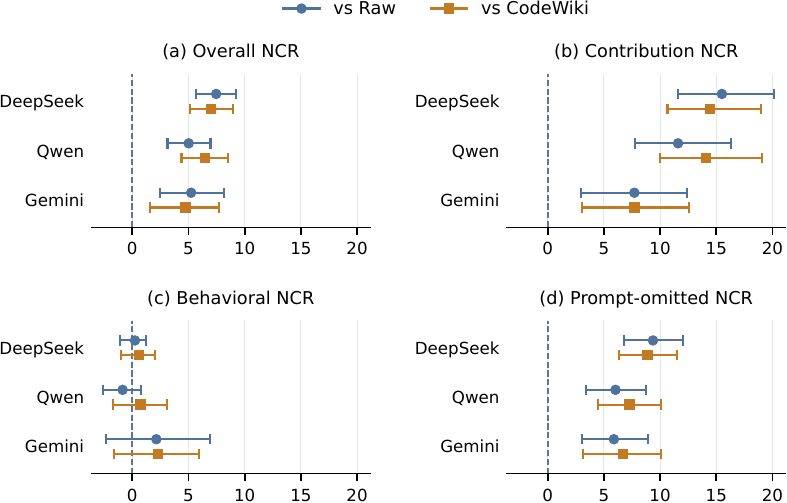}
    \caption{Mean paired NCR differences (\app minus baseline, in percentage points), with pointwise BCa 95\% intervals over 118--121 tasks decidable in both conditions.}
    \Description{Four panels show paired NCR differences in percentage points and 95 percent confidence intervals for Overall, Contribution, Behavioral, and Prompt-omitted NCR across DeepSeek, Qwen, and Gemini. Circles compare RepoNorm with Raw; squares compare it with CodeWiki. Intervals for Overall, Contribution, and Prompt-omitted NCR are entirely positive. All intervals for Behavioral NCR include zero.}
    \label{fig:eval-paired-changes}
\end{figure}

\textbf{Contribution and behavioral compliance.} The larger gains occur in Contribution NCR: \app improves over Raw by 31.64--45.44\% and over CodeWiki by 29.86--48.03\%, relative to each baseline. Behavioral NCR shows smaller, mixed changes. \app is slightly below Raw with DeepSeek and Qwen and above Raw with Gemini; it exceeds CodeWiki with all three models. The paired 95\% intervals for Behavioral NCR include zero in all six comparisons (Figure~\ref{fig:eval-paired-changes}). By comparison, the intervals for Overall, Contribution, and Prompt-omitted NCR are entirely positive in all 18 comparisons.

\textbf{Prompt-omitted compliance.} Prompt-omitted NCR reaches 61.28\%, 55.81\%, and 49.68\% with \app for DeepSeek, Qwen, and Gemini, respectively, yielding relative gains of 11.34--17.69\% over Raw and 15.00--16.93\% over CodeWiki. The gains extend to norms not explicitly stated in task descriptions, consistent with repository norm acquisition supplementing task instructions.

\textbf{Interpretation.} A plausible explanation lies in how the resources present repository knowledge. CodeWiki documentation can explain implementation details and provide development advice, leaving the agent to identify which statements impose obligations on its changes. \app organizes guidance around norms with explicit scope, conditions, and supporting evidence. This organization may help agents identify contribution work beyond implementing the feature, consistent with the larger gains in Contribution NCR.

\paragraph{\textbf{Sensitivity to benchmark revisions.}}
We assess whether removing implementation guidance during benchmark revision could account for \app's compliance gains.

\textbf{Design.} We compare the original and final descriptions for the 670 Prompt-omitted task--norm pairs. Of these, 651 already lacked reminders in the original descriptions, while partial reminders were removed for 16 pairs. The remaining three pairs are excluded from this analysis because their correspondence is unclear. Excluding all 14 tasks with confirmed reminder removals leaves 107 tasks.

\textbf{Results.} On the 651 originally unmentioned associations, \app's NCR is 10.89--18.37\% higher than Raw and 14.99--17.88\% higher than CodeWiki across the three models, relative to each baseline. On the 107-task subset, \app's Overall NCR remains 7.85--10.63\% higher than Raw and 8.77--11.18\% higher than CodeWiki, relative to each baseline. The paired 95\% intervals remain entirely positive in both analyses. The gains therefore persist beyond the associations and tasks containing reminder removals.

\paragraph{\textbf{Reliability of norm evaluation.}}
We assess reference annotations, norm-check accuracy, and sensitivity to the documentation proxy.

\textbf{Design.} We review fixed stratified samples of 42 reference norms and 84 task--norm associations, selecting three norms and six associations per repository version. Sampling accounts for norm categories and, for associations, the existing prompt-explicitness annotations. We assess norm validity and evidential support, task applicability, and prompt explicitness against repository evidence and task descriptions.

During development, we constructed 455 patches: 114 expected to comply with the applicable norms and 341 designed to violate specific norms. Checking the compliant patches against their applicable norms and the violating variants against their targeted norms yielded 1,351 norm checks. Documentation checks use a change signal as a proxy for delivery, which does not guarantee content correctness. We assess sensitivity to this limitation by excluding the same 59 documentation task--norm pairs from every condition and model.

\textbf{Results.} All sampled norms are valid, with full evidential support for 41 and partial support for one. All sampled associations are applicable; their task descriptions fully state the norm in 11 cases, partly state it in 20, and omit it in 53.

All 1,010 checks with expected compliance passed, with no false rejections. The checks detected 302 of the 341 expected violations. The remaining 39 were all misleading-documentation variants, so every tested violation outside this specific boundary was detected. These misses are expected given the documentation proxy.

\app's Overall NCR remains 8.57--11.53\% higher than Raw and 8.59--11.85\% higher than CodeWiki, relative to each baseline. Contribution NCR under the same exclusion remains 31.21--55.84\% higher than Raw and 33.77--63.31\% higher than CodeWiki. All paired 95\% intervals remain entirely positive. The advantages in these two measures persist without these documentation checks.

\begin{rqfindings}{1}
\app improves Overall, Contribution, and Prompt-omitted NCR over both baselines with every evaluated model. Gains are larger for contribution requirements less directly tied to functional correctness; changes in Behavioral NCR are smaller and uncertain.
\end{rqfindings}

\subsection{RQ2: Functional Correctness}
\label{sec:eval-correctness}

\textit{Does \app improve the functional correctness of coding-agent solutions?}

\paragraph{\textbf{Design.}}
We measure functional success as the fraction of all 121 tasks whose solutions pass all functional tests~\cite{jimenez2024swebenchlanguagemodelsresolve}. \textit{Joint success} requires functional success and passing every applicable norm check, with at least one applicable norm. Table~\ref{tab:eval-correctness} reports overall results for each model.

We compare \app with each baseline using two-sided exact McNemar tests on the paired binary task outcomes. Each test contrasts tasks successful with \app but not the baseline against tasks successful with the baseline but not \app. We report raw $p$ values and apply Holm correction across all 12 functional and joint comparisons (three models, two baselines, two outcomes), using a 0.05 threshold.

% Generated by artifacts/evaluation/build_statistics.py.
\newsavebox{\evalCorrectnessRates}
\newsavebox{\evalCorrectnessTests}
\begin{table}[htbp]
\centering
\caption{Overall functional and joint success, with paired tests.}
\label{tab:eval-correctness}
% \small
\sbox{\evalCorrectnessRates}{%
\resizebox{0.95\width}{!}{%
\begin{tabularx}{\linewidth}{@{}cl*{5}{>{\centering\arraybackslash}X}@{}}
\toprule
\multirow{2}{*}{Model} & \multirow{2}{*}{Outcome} & Raw & \multicolumn{2}{c}{CodeWiki} & \multicolumn{2}{c}{\textbf{\app}} \\
\cmidrule(lr){3-3}\cmidrule(lr){4-5}\cmidrule(lr){6-7}
 &  & Rate (\%) & Rate (\%) & $\Delta$ (pp) & Rate (\%) & $\Delta$ (pp) \\
\midrule
\multirow{2}{*}{DeepSeek} & Functional & 59.50 & 66.94 & \evalup{7.44} & \textbf{68.60} & \evalup{9.09} \\
 & Joint & 3.31 & 4.13 & \evalup{0.83} & \textbf{8.26} & \evalup{4.96} \\
\midrule
\multirow{2}{*}{Qwen} & Functional & 44.63 & 44.63 & \evalflat{0.00} & \textbf{52.07} & \evalup{7.44} \\
 & Joint & 2.48 & 0.83 & \evaldown{1.65} & \textbf{4.96} & \evalup{2.48} \\
\midrule
\multirow{2}{*}{Gemini} & Functional & 30.58 & \textbf{37.19} & \evalup{6.61} & 36.36 & \evalup{5.79} \\
 & Joint & 0.83 & 0.83 & \evalflat{0.00} & \textbf{1.65} & \evalup{0.83} \\
\bottomrule
\end{tabularx}%
}%
}
\usebox{\evalCorrectnessRates}
\par\medskip
% \small
\sbox{\evalCorrectnessTests}{%
\resizebox{0.95\width}{!}{%
\begin{tabular*}{\linewidth}{@{\extracolsep{\fill}}llrr@{}}
\toprule
Model & Outcome & \app vs Raw & \app vs CodeWiki \\
\midrule
\multirow{2}{*}{DeepSeek} & Functional & 0.052 (0.575) & 0.839 (1.000) \\
 & Joint & 0.031 (0.375) & 0.125 (1.000) \\
\midrule
\multirow{2}{*}{Qwen} & Functional & 0.136 (1.000) & 0.122 (1.000) \\
 & Joint & 0.250 (1.000) & 0.063 (0.625) \\
\midrule
\multirow{2}{*}{Gemini} & Functional & 0.248 (1.000) & 1.000 (1.000) \\
 & Joint & 1.000 (1.000) & 1.000 (1.000) \\
\bottomrule
\end{tabular*}%
}%
}
\usebox{\evalCorrectnessTests}
\par\smallskip
\begin{minipage}{\maxof{\wd\evalCorrectnessRates}{\wd\evalCorrectnessTests}}
\footnotesize
\raggedright
All rates use 121 tasks. $\Delta$: change from Raw in percentage points (pp), computed before rounding. Bold: highest rate per model and outcome. Below: exact McNemar $p$ values (Holm-adjusted in parentheses).
\end{minipage}
\end{table}

\paragraph{\textbf{Results.}}
We report functional and joint success alongside their paired statistical comparisons.

\textbf{Functional success.} \app's observed functional success rates are 9.09, 7.44, and 5.79 percentage points higher than with Raw for DeepSeek, Qwen, and Gemini, respectively. Against CodeWiki, the rates are 1.65 and 7.44 percentage points higher for DeepSeek and Qwen, respectively, and 0.83 percentage points lower for Gemini. None of the six functional-success comparisons reaches the 0.05 threshold.

\textbf{Joint success.} \app has the highest observed joint success for all three models, with ten, six, and two successful tasks, compared with four, three, and one for Raw and five, one, and one for CodeWiki. None of the 12 binary comparisons reaches the threshold after correction. Joint success remains low across all models and conditions, indicating that functional success often leaves some repository requirements unmet.

\textbf{Interpretation.} General repository context and explicit norm guidance may both assist task completion. The observed functional differences are insufficient to establish an improvement under these paired tests.

\begin{rqfindings}{2}
\app has higher observed functional success than Raw and higher observed joint success than both baselines, but the paired binary comparisons do not establish an improvement after multiplicity correction. Joint success remains uncommon.
\end{rqfindings}

\subsection{RQ3: Norm Quality}
\label{sec:eval-quality}

\textit{How accurate are the acquired norms, and how well do they cover task-relevant repository norms?}

\paragraph{\textbf{Design.}}
We assess precision, reference-norm coverage, and assessment reliability for both views.

\textbf{Precision.} We assess both the Accepted-Only View and the automated Accepted-after-Review View described in Section~\ref{sec:approach-delivery}. The two views contain 1,990 and 6,277 readable norm entries across 14 packages, respectively. For the latter, an automated reviewer (GPT-5.6-Sol-high) accepts additional candidates in a separate review session. We assess separate fixed samples of 300 and 600 entries, allocating samples approximately in proportion to each package's population and selecting entries by a seeded hash ordering within each package. GPT-6-Astra assesses norm correctness. Precision is the number of entries judged correct divided by the number assessed. An entry is judged correct only when all of its fields are accurate.

\textbf{Coverage of task-relevant reference norms.} We assess whether each of the 160 reference norms is covered by each view, allowing multiple entries to jointly support a norm. \textit{Full coverage} means that a reference norm is covered in its entirety. \textit{Partial coverage} means that only part of it is covered, such as an individual clause. The remaining norms are \textit{uncovered}. With $G$ reference norms, $F$ fully covered norms, and $P$ partially covered norms, we report:
\begin{equation}
    \mathrm{Coverage}_{\mathrm{full}}=\frac{F}{G},
    \qquad
    \mathrm{Coverage}_{\mathrm{full+partial}}=\frac{F+P}{G}.
    \label{eq:gold-coverage}
\end{equation}
Both measures concern this task-relevant reference set.

\textbf{Assessment reliability.} To assess the reliability of GPT-6-Astra's judgments, we select a fixed subsample of 140 entries, drawing five entries per package from each view's assessed sample. Two human reviewers independently reassess these entries against repository evidence, with view labels, GPT-6-Astra's judgments, and the other reviewer's judgments withheld. We then compare the reviewers' shared judgments with GPT-6-Astra's judgments.

\newsavebox{\evalQualityTable}
\begin{table}[htbp]
\centering
\caption{Overall norm quality before and after review.}
\label{tab:eval-quality}
% \small
\sbox{\evalQualityTable}{%
\resizebox{0.95\width}{!}{%
\begin{tabular*}{\linewidth}{@{\extracolsep{\fill}}lrrr@{}}
\toprule
Norm view & Precision & Full coverage & Full+partial coverage \\
\midrule
Accepted-Only & 258/300 (86.00) & 35/160 (21.88) & 63/160 (39.38) \\
Accepted-after-Review & 451/600 (75.17) & 84/160 (52.50) & 125/160 (78.13) \\
\bottomrule
\end{tabular*}%
}%
}
\usebox{\evalQualityTable}
\par\smallskip
\begin{minipage}{\wd\evalQualityTable}
\footnotesize
\raggedright
Cells show numerator/denominator (\%). Precision uses separate samples; coverage uses the same norms for both views.
\end{minipage}
\end{table}

\paragraph{\textbf{Results.}}
Table~\ref{tab:eval-quality} summarizes overall precision and reference-norm coverage for the two views.

\textbf{Precision and coverage.} The Accepted-Only View achieves 86.00\% sampled precision, with 21.88\% full and 39.38\% full+partial coverage of task-relevant reference norms. The Accepted-after-Review View expands these coverage rates to 52.50\% and 78.13\%, respectively, with lower sampled precision of 75.17\%.

\textbf{Assessment reliability.} The two reviewers agree on 126 of the 140 entries (90.00\%). Among these 126 entries, their shared judgments match GPT-6-Astra's assessments on 114 entries (90.48\%). This agreement suggests that the model's correctness judgments are consistent with the reviewers' assessments, supporting the reliability of our assessment on the sampled entries.

\textbf{Interpretation.} Post-extraction review admits additional candidates withheld by conservative verification. The resulting view has higher reference-norm coverage and lower sampled precision. Our criterion counts an entry as incorrect if any field is inaccurate, even when its task-relevant guidance remains valid. Consequently, some entries classified as incorrect may still contain task-relevant guidance.

\begin{rqfindings}{3}
The Accepted-Only View has higher sampled precision but limited coverage of task-relevant reference norms. The Accepted-after-Review View expands both full and full+partial coverage, with lower sampled precision.
\end{rqfindings}

\subsection{RQ4: Cost Analysis}
\label{sec:eval-cost}

\textit{What are the time and token costs of norm acquisition and guided coding?}

\paragraph{\textbf{Design.}}
We measure recorded input and output tokens and cumulative execution time for norm acquisition and coding. Acquisition is counted once for each of the 14 Accepted-Only packages; coding costs are summed over all 121 tasks for each model and condition. Token totals reflect recorded usage; some requests lack token counts.

\newsavebox{\evalCostTable}
\begin{table}[htbp]
\centering
\caption{Recorded token usage and cumulative execution time for norm acquisition and coding.}
\label{tab:eval-cost}
\small
\sbox{\evalCostTable}{%
\resizebox{0.95\width}{!}{%
\begin{tabular}{@{}lrrr@{}}
\toprule
Stage & \makecell[r]{Input tokens (M)} & \makecell[r]{Output tokens (M)} & \makecell[r]{Cumulative execution time (h)} \\
\midrule
Norm acquisition & 134.52 & 7.67 & 2.14 \\
\midrule
Coding with Raw & 346.45 & 10.10 & 53.56 \\
\addlinespace
Coding with CodeWiki & 391.73 & 10.89 & 63.72 \\
\addlinespace
Coding with \app & 463.83 & 12.70 & 67.12 \\
\bottomrule
\end{tabular}%
}%
}
\usebox{\evalCostTable}
\par\smallskip
\begin{minipage}{\wd\evalCostTable}
\footnotesize
\raggedright
Acquisition: 14 packages. Coding: 121 tasks $\times$ 3 models. M: million tokens. Token records are incomplete.
\end{minipage}
\end{table}

\paragraph{\textbf{Results.}}
Generating the 14 reusable Accepted-Only norm packages took an average of 9.18 minutes per package, with a median of 8.86 minutes. Their one-time cumulative acquisition time was 2.14 hours, equivalent to 3.2\% of the 67.12 hours of cumulative coding-agent execution time with \app (Table~\ref{tab:eval-cost}). Acquisition accounts for 23.0\% of the combined recorded token usage for acquisition and coding with \app. Across the three coding models, CodeWiki and \app use 12.92\% and 33.65\% more recorded coding tokens than Raw, respectively, with 18.98\% and 25.33\% more cumulative execution time. With Gemini, \app uses less cumulative coding time than CodeWiki. Reading and applying the additional documentation or norm packages can expand the model context and require further agent actions, which may explain these higher aggregate costs.

\begin{rqfindings}{4}
Norm packages take 9.18 minutes each on average to acquire and can be reused across tasks and models. Coding with CodeWiki or \app uses more recorded tokens and cumulative execution time than Raw in aggregate.
\end{rqfindings}

\section{Threats to Validity}
\label{sec:threats}
\textbf{Measurement validity.}
The authors constructed the reference norms with model assistance without consulting \app's packages. These norms may be incomplete and reflect author judgment, so coverage measures recovery of this task-relevant reference set. Precision sampling may miss rare errors or imperfectly represent the output population. Incomplete token records may bias cost comparisons if missingness differs across conditions.

\textbf{Statistical inference.}
Tasks share repository versions, so task-level bootstrap intervals and paired tests may understate uncertainty and weaken inference about new repositories. One recorded execution per model--condition--task does not measure variation across repeated runs. Category-group analyses are exploratory, and NCR intervals are pointwise rather than simultaneous.

\textbf{External validity.}
Our evaluation covers six relatively small repositories, three coding models, and one agent harness. Source analysis currently supports Python. Effectiveness and cost for other languages, larger repositories, and other agent configurations remain untested.

\textbf{Practical limitations.}
Our evaluation does not isolate component contributions or establish optimal parameter settings. Fixed analysis budgets may miss norms or counterexamples as repositories grow. Repository changes can make packages stale. Cache hits avoid repeated model calls, but incremental updating is not yet sufficiently optimized for repository evolution. Frequent or extensive changes require ongoing investment of time and resources. Benefits also depend on agents selecting and applying relevant norms, which they may overlook among many entries~\cite{liu-etal-2024-lost,jaroslawicz2025instructionsllmsfollowonce}.

\section{Related Work}
\label{sec:related}
\textbf{Mining conventions and specifications.}
Prior work recovers programming conventions and constraints from source code~\cite{allamanis2014miningidiomssourcecode,singleton2019inferringconcisespecificationsapis}, execution traces~\cite{kang2021adversarialspecificationmining}, and documentation~\cite{grent2021automaticallyidentifyingparameterconstraints,naziri2026testingdeeplearninglibraries}. NATURALIZE learns naming and formatting conventions with statistical language models~\cite{allamanis2014naturalize}. Other methods enable specification mining from abstracted symbolic traces~\cite{henkel2019enablingopenworldspecificationmining} or extract input constraints from API documentation~\cite{xie2024docterdocumentationguidedfuzzing}. SpecGen refines formal specifications through verifier feedback~\cite{ma2025specgenautomatedgenerationformal}, while RBCTest mines response-body constraints from OpenAPI descriptions and checks generated validators against specification examples~\cite{huynh2026rbctestleveragingllmsverify}. \app acquires norms governing implementation, tests, and documentation, checking applicability against evidence and counterexamples.

\textbf{Repository context during coding.}
Cross-file context supports repository-level code completion~\cite{liu2023repobenchbenchmarkingrepositorylevelcode,ding2023crosscodeevaldiversemultilingualbenchmark}; RepoCoder alternates retrieval and generation~\cite{zhang-etal-2023-repocoder}, while Repoformer selectively retrieves cross-file context~\cite{pmlr-v235-wu24a}. For repository modification, SWE-agent supports navigation, editing, and execution feedback~\cite{yang2024sweagent}. Other systems likewise acquire and use task-specific context~\cite{zhang2024autocoderoverautonomousprogramimprovement,xia2024agentlessdemystifyingllmbasedsoftware,ouyang2025repographenhancingaisoftware}. \app prepares norms before a task is specified, complementing the consuming agent's context selection and coding loop.

\textbf{Repository documentation and reusable guidance.}
RepoAgent generates documentation from code structure and reference relationships~\cite{luo-etal-2024-repoagent}; CodeWiki uses hierarchical decomposition and synthesis~\cite{hoang-etal-2026-codewiki}. Instruction files provide reusable guidance~\cite{chatlatanagulchai2026agentreadmesempiricalstudy,lulla2026impactagentsmdfilesefficiency}. Gloaguen et al. evaluate developer-written and LLM-generated files, finding that agents generally follow instructions without consistent task-resolution gains in the evaluated settings~\cite{gloaguen2026agentsmd}. Probe-and-Refine revises reusable guidance using feedback from synthetic bug-fix tasks~\cite{shepard2026probeandrefinetuningrepositoryguidance}. \app checks norms' content and applicability against repository evidence, attaching support, conditions, and exceptions to guidance.

\textbf{Checking project-specific compliance.}
SWE-Shield extracts design constraints from pull-request reviews and assesses issue-linked patches with an LLM verifier separately from functional tests~\cite{yu2026doespassratetell}. OctoBench scores recorded agent trajectories against task-specific instruction checklists~\cite{ding-etal-2026-octobench}. Our evaluation separates norm precision and reference coverage from downstream compliance and functional success.

\section{Conclusion}
\label{sec:conclusion}
\app acquires repository norms as reusable guidance whose content and applicability are checked against repository evidence. On 121 RepoNormBench tasks across three coding models, \app achieves relative improvements over Raw of 7.42--10.77\% in Overall NCR, 31.64--45.44\% in Contribution NCR, and 11.34--17.69\% in Prompt-omitted NCR. It also exceeds CodeWiki on all three measures for every model. Observed functional success rates are 5.79--9.09 percentage points higher than with Raw, although paired functional and joint-success comparisons do not establish a statistically significant improvement after multiplicity correction. Joint success remains uncommon. Under its default configuration, \app achieves 86.00\% sampled precision in norm acquisition. These findings show that contribution completeness remains a distinct challenge even when functional tests pass. They support making repository obligations explicit in agent guidance and evaluating their satisfaction alongside functional correctness.

\section{Data Availability}

To facilitate replication, we have released our data and code at: \url{https://github.com/SYSUSELab/RepoNorm}.

\bibliographystyle{ACM-Reference-Format}
\bibliography{references}

\end{document}